\documentclass[prl,nofootinbib,superscriptaddress,twocolumn]{revtex4}

\usepackage{amssymb}
\usepackage{amsmath}
\usepackage{graphicx}
\usepackage{longtable}
\usepackage{verbatim}
\usepackage{amsfonts}

\newcommand{\be}{\begin{equation}}
\newcommand{\ee}{\end{equation}}
\newcommand{\bea}{\begin{eqnarray}}
\newcommand{\eea}{\end{eqnarray}}
\newcommand{\beq}{\begin{equation}}
\newcommand{\eeq}{\end{equation}}
\def\beqa{\begin{eqnarray}}

\def\eeqa{\end{eqnarray}}

\def\lsim{\mathrel{\rlap{\lower4pt\hbox{\hskip0.5pt$\sim$}}
    \raise1pt\hbox{$<$}}}         
\def\gsim{\mathrel{\rlap{\lower4pt\hbox{\hskip0.5pt$\sim$}}
    \raise1pt\hbox{$>$}}}         

\begin{document}

\vspace*{-30mm}

\title{Remarks about the Tensor Mode Detection by the  BICEP2 Collaboration \\
\vskip 0.1cm
 and the Super-Planckian Excursions of the Inflaton Field}
\vskip 0.3cm
\author{A. Kehagias }
\author{A. Riotto}
\address{D\'epartement de Physique Th\'eorique and Centre for
  Astroparticle Physics (CAP), Universit\'e de Gen\`eve, 24 quai E. Ansermet, CH-1211 Geneva, Switzerland
}

\date{\today}

%
%

\begin{abstract}
\noindent
The recent detection by the BICEP2 collaboration  of a high level of tensor modes  has relevant implications  which we briefly discuss in this short note.  In particular,  the large angle CMB B-mode polarisation seems to imply problematic super-Planckian excursions of the inflaton field. We provide some comments about this point and in particular we stress a natural resolution to it: given our current (and probably future) observational ignorance about the true source of the scalar perturbations, one should abandon the  theoretical prejudice that they are associated to the inflaton fluctuations. 

\end{abstract}

\maketitle

\noindent
\noindent
\noindent
Inflation \cite{guth81,lrreview} has  become the dominant 
paradigm for understanding the 
initial conditions for structure formation and for Cosmic
Microwave Background (CMB) anisotropy. In the
inflationary picture, primordial density and gravity-wave fluctuations are
created from quantum fluctuations ``redshifted'' out of the horizon during an
early period of superluminal expansion of the universe, where they
are ``frozen'' \cite{muk81,bardeen83}. 

The recent measurement of the tensor
modes from large angle CMB B-mode polarisation by
BICEP2 \cite{bicep}, implying  a tensor-to-scalar ratio 

\be 
r=0.2^{+0.07}_{-0.05},
\ee
has put inflation on a ground which is  firmer than ever.
Indeed, the generation of gravity-wave fluctuations 
is a generic prediction of an accelerated  de Sitter expansion of the universe.
The tensor modes 
may be viewed as  ripples of spacetime around the  background metric

\be
g_{\mu\nu}={\rm d}t^2-a^2(t)\left(\delta_{ij}+h_{ij}\right)
{\rm d}x^i {\rm d}x^j,
\ee
 where $a$ is the scale factor and $t$ is the cosmic time. The  
tensor $h_{ij}$ is traceless and transverse and has two  polarizations,
$\lambda=\pm$.
Since gravity-wave fluctuations are (nearly)
frozen on super-Hubble scales,
a way of characterizing them is to compute
their spectrum on scales larger than the Hubble radius during inflation. 
The 
power spectrum of gravity-wave  modes turns out to be

\begin{equation}
{\cal P}_{T}(k)=\frac{8}{M_p^2}\left(\frac{H_*}{2\pi}\right)^2
\left(\frac{k}{aH_*}\right)^{-2\epsilon},
\end{equation}
where $M_p=(8\pi G)^{-1/2}\simeq 2.4\times 10^{18}$ GeV 
is the Planck scale. Here
$\epsilon=(\dot{\phi}^2/2M_p^2 H_*^2)$ is a standard  slow-roll parameter and $H_*=\dot a/a$
indicates the Hubble rate during inflation. 

On the other hand, 
the power spectrum of curvature
perturbations in slow-roll inflationary models is given by

\begin{equation}
{\cal P}_{\zeta}(k)=
\frac{1}{2 M_p^2\epsilon}\left(\frac{H_*}{2\pi}\right)^2
\left(\frac{k}{aH_*}\right)^{n_\zeta-1},
\label{pscalar}
\end{equation}
where $n_{\zeta}\simeq 1$ is the spectral index. Since the fractional changes
of the power spectra with  scales are  much smaller than unity, one can safely 
consider the power spectra as roughly constant on the  scales relevant for the
CMB anisotropy and define a tensor-to-scalar amplitude ratio 

\begin{equation}
\label{q}
r=\frac{{\cal P}_{T}}{16\,{\cal P}_{\zeta}}=\epsilon\, .
\end{equation}
The recent  BICEP2 dataset  allows
to extract the value of the Hubble rate during inflation to be

\begin{equation}
\label{bicepH}
H_*\simeq 1.1\times 10^{14}\,{\rm GeV}, 
\end{equation}
corresponding to an energy scale during inflation $V^{1/4}$ of about $2\times 10^{16}\,{\rm GeV}$, astonishingly
closed to the scale of grand-unification in the minimal supersymmetric extension of the standard model
of weak interactions.

Let us now pause for a moment and summarize what we really know about the properties
of the scalar and tensor perturbations generated during inflation:

\begin{itemize}

\item First of all, the recent Planck data \cite{ade} tell us that the scalar  perturbations have an almost  scale-invariant
spectrum and 
are of the adiabatic type, but we do not know the real source of the scalar perturbations. This point will be relevant below.

\item The scalar perturbations are nearly-Gaussian or, in any case, the level of non-Gaussianity, parametrized by the non-linear parameter $f_{\rm NL}$, is 
 severely constrained \cite{ng}.

\item The energy scale of inflation is approximately the grand-unification  scale.
\end{itemize}
These are consequences of the observational facts which nobody can dispute. 
Here are some implications one can draw:

\begin{itemize}
\item In the sudden reheating approximation, the  maximum reheating temperature after inflation of about

\begin{eqnarray}
T_{\rm RH}&=&\left(\frac{30\, V}{\pi^2 g_*(T_{\rm RH})}\right)^{1/4}\nonumber\\&\simeq&5.6\left(\frac{10^3}{g_*(T_{\rm RH})}\right)^{1/4}\times 10^{15}\,{\rm GeV}.
\end{eqnarray}
The true reheating temperature is likely to be smaller; one should also remember that at temperatures larger than about $2.4\times 10^{14}$ GeV the universe is not in thermal equilibrium and one may not define a temperature \cite{enq}.

\item The Standard Model Higgs field $h$ needs to be non-trivially coupled either  to the inflaton field or to gravity. Indeed, 
for a Higgs mass in the range
$(124-126)$ GeV, and for the current central values of the top mass and strong coupling
constant, the Higgs potential develops an instability around
$10^{11}$ GeV, see for instance Ref. \cite{noi1}. As this instability scale is much smaller than $H_*$,   the classical value of the Higgs field will be easily pushed
above the instability point by its fluctuations during inflation \cite{noi2}. This can be avoided 
by either coupling the Higgs field to the Ricci scalar, $\xi R h^\dagger h$ with $\xi\gsim 10^{-1}$, or to the inflaton itself in order to suppress the Higgs fluctuations during inflation. 

\item Similar  remarks can be drawn for the case in which supersymmetry
is a (broken) symmetry of nature. Indeed, there are many flat directions
in the field space of low-energy supersymmetric models. It may happen that
some combination of the squark and/or slepton mass squared
parameters get negative at some scale below the
Planck scale when running through the renormalization
group equations from the weak scale up. This
may happen if the sfermion masses are lighter than the
gaugino masses, leading either to the appearance of unacceptable
color/charge breaking unbounded from below directions
in the effective potential for the squark and/or slepton
fields. The instability case can be again smaller than $H_*$, posing a threat
during inflation due to the large fluctuations of the sfermion fields \cite{me}. Again, one is led to conclude that low-energy supersymmetric partners must be coupled to the inflation field.

\end{itemize}
Furthermore, and maybe more interestingly, the recent
detection of a high level of tensor modes have generated a lot of surprise based on the following argument due to Lyth \cite{lythgrav}.

If the scalar perturbations are ascribable to only one scalar degree of freedom, the inflaton field itself (this is not a gauge-invariant statement), then
the slow-roll paradigm gives, using
the definition of $\epsilon$ and 
Eq. (\ref{q}), 

\begin{equation}
\label{k}
\frac{1}{M_p}\left|\frac{{\rm d}\phi}{{\rm d}N}\right|=\sqrt{2}\,r^{1/2},
\end{equation}
where ${\rm d}\phi$ is the change in the inflaton field in ${\rm d}N=H{\rm d}t\simeq
{\rm d}\,{\rm ln}\, a$ Hubble times. While the scales corresponding to
the relevant multipoles in the CMB anisotropy are living the Hubble radius
$\Delta N\simeq 4.6$ and therefore the field variation
is 

\be
\frac{\Delta\phi}{M_p}\simeq \left(\frac{r}{2\times 10^{-2}}\right)^{1/2}.
\ee
 This is a minimum
estimate because inflation continues for some number $N$ of $e$-folds
of order of 50.
The detection of   gravitational waves requires in general
variation of the inflaton field of the order of the Planck scale \cite{lythgrav}. 

This conclusion is considered to be a problem as  
 slow-roll models of inflation are generically
based on four-dimensional field theories, possibly involving supergravity,
where higher-order operators with powers of $(\phi/M_p)$ are disregarded. This
assumption is justified only if the inflaton variation is small
compared to the Planck scale. It is therefore difficult to 
construct a satisfactory model of inflation firmly rooted  
in modern particle theories having possibly supersymmetry as a
crucial ingredient and with large variation of the inflaton field.

It is more than fair to say that, based on this argument, there was  a strong theoretical
prejudice against the likelihood of observation of gravity-waves.
So, now that a high level of tensor modes have been observed, where do we stand? Do we still believe that
Planckian excursions of the inflaton field is a threat? 

There are at least three arguments we may offer in favour of a more relaxed attitude.

The first one is in fact quite simple: it is a theoretical prejudice that 
the scalar perturbations come from the inflaton field. 

The sad reality  is that  we have
no idea what is the real  source of the scalar perturbations during inflation. Even worse, in the absence of a detection of
a large non-Gaussianity, we will probably never know. 

The problem of having  large excursion of the inflaton field arises
only if the scalar perturbations are generated by the inflaton itself, which is the origin of the relation
(\ref{k}).  
Despite the simplicity of the inflationary paradigm, the mechanism
by which  cosmological adiabatic perturbations are generated  is not
yet established. 
It is conceivable that 
the total curvature perturbation $\zeta$  is not a constant (in time) on super-Hubble scales, 
but on the contrary  changes on arbitrarily large scales due to a non-adiabatic
pressure perturbation    which may be 
present  extra (other than the inflaton) degrees of freedom are present.
While the entropy
perturbations evolve independently of the curvature perturbation on
large scales,  the evolution of the large-scale curvature is
sourced by the entropy perturbation $\delta S$

\be
\dot\zeta\sim \delta S.
\ee
A realization of this mechanism 
 is represented, for instance,  by the curvaton 
mechanism
\cite{curvaton1,LW,curvaton3,LUW} where the final curvature perturbations
are produced from an initial isocurvature perturbation associated to the
quantum fluctuations of a light scalar field (other than the inflaton), 
the curvaton, whose energy density is negligible during inflation. The 
curvaton isocurvature perturbations are transformed into adiabatic
ones when the curvaton decays into radiation much after the end 
of inflation. 
Other mechanisms for the generation of cosmological
perturbations have  been proposed, for instance the modulated decay scenario \cite{gamma1,gamma2,gamma3}, 
where    super-Hubble spatial
fluctuations in the decay rate of the inflaton field
are induced during inflation, causing  adiabatic perturbations
in  the final reheating temperature
in different regions of the universe. Also, the dominant contribution to the primordial curvature perturbation may be generated at the end of inflation \cite{end1,end2}.

Consider,
for instance, the  simplest curvaton scenario \cite{LW},  being $\sigma$ the curvaton field.
During inflation, the curvaton energy density is 
negligible and isocurvature perturbations with
a flat spectrum are produced in the curvaton field $\sigma$,
$\langle\delta\sigma^2\rangle^{\frac{1}{2}} = (H_*/2\pi)$, 
where $\sigma_*$ is the value of the curvaton field during inflation.
After the end
of inflation, 
the curvaton field oscillates during some radiation-dominated era,
causing its energy density to grow and 
thereby converting the initial isocurvature into curvature 
perturbation.
After the curvaton decays $\zeta$  becomes constant. In 
the approximation that the curvaton decays instantly
it is then given by $
\zeta \simeq (2\gamma/3) \left(\delta \sigma/\sigma \right)_*$, 
where $\gamma\equiv (\rho_\sigma/\rho)_{D}$ 
and the subscript $D$ denotes the epoch of decay. The corresponding spectrum
is \cite{LW}

\begin{equation}
\label{spectrum}
{\cal P}_\zeta^{\frac{1}{2}}\simeq
\frac{2\gamma}{3}  \left(\frac{H_*}{2\pi \sigma_*}\right).
\end{equation}
Since the amplitude of curvature perturbation ${\cal P}_\zeta^{1/2}$ 
must
match the observed  value $5\times
10^{-5}$, from Eq. (\ref{spectrum}) one infers that

\begin{equation}
\sigma_*\simeq 2\,\gamma\,\times 10^3\, H_*.
\end{equation}
For $10^{-1}\lsim \gamma\lsim 1$, the corresponding level of non-Gaussianity is such  that $-5/4\lsim f_{\rm NL}\lsim 5/4r$ \cite{reviewng}.
Since a  level of (local) non-Gaussianity compatible with the present Planck data  is $f_{\rm NL}\lsim 10$ \cite{reviewng},  we conclude that

\begin{equation}
\label{17}
\left(2\times 10^{16}\, {\rm GeV}\lsim \sigma_*\lsim 2\times 10^{17}\, {\rm GeV}\right).
\end{equation}
This is is comfortably below the Planckian scale. Of course, we are working under the assumption that the 
the curvature perturbations
of the inflaton field are  suppressed. This may happen, for instance, 
if 
the inflaton field  is well anchored at the false vacuum driving inflation 
with a mass much $m_\phi$ larger than the Hubble rate during inflation.
Suppose indeed that the inflaton potential is of the form

\begin{equation}
V(\phi)= V_0+\frac{m_\phi^2}{2}\phi^2+\cdots,
\end{equation}
where  $\phi=0$ is the location of the minimum around which
$m^2_\phi\gg H^2_*\simeq V_0/M_p$.
Under these circumstances, slow-roll conditions are badly violated
since $\eta=(m_\phi^2/3 H_*^2)\gsim 1$
and 
the fluctuations of the inflaton 
field on super-Hubble scales read 

\begin{equation}
{\cal P}_{\delta \phi}(k)=
\Big(\frac{H_*}{2\pi}\Big)^2 \Big(\frac{k}{aH_*}\Big)^3\, e^{-2 m_\phi^2/H_*^2}\, .
\end{equation}
The resulting power spectrum is suppressed \cite{alberto}. 
This scenario just needs an extra  degree of freedom which act as a clock to remove the inflaton from its false vacuum, thus ending inflation. A red spectrum for the curvature perturbations can be easily obtained by supposing that the curvaton field
during inflation is slowly rolling to its true vacuum from the top to its potential, such that its effective mass squared $m^2$ is negative and $n_\zeta=1+(2 m^2/3 H_*^2)\simeq 0.96$.
%
%
Similar considerations are  obtained in the modulated decay scenario where the inflaton decay rate $\Gamma$ depends on a light field $\sigma$ quadratically, $\Gamma\sim \sigma^2$. The corresponding power spectrum   reads \cite{reviewng}

\begin{equation}
\label{spectrum1}
{\cal P}_\zeta^{\frac{1}{2}}\simeq\frac{1}{6}\frac{{\rm d}\ln \Gamma}{{\rm d}\sigma}\langle\delta\sigma^2\rangle^{\frac{1}{2}} =
\frac{1}{3}  \left(\frac{H_*}{2\pi \sigma_*}\right) 
\end{equation}
and non-Gaussianity parameter is small, $f_{\rm NL}\simeq 5/2$.

Suppose though that we insist in taking a minimalistic approach and restrict ourselves to the standard scenario where the scalar perturbations are due to the inflaton itself. 

The second argument of why we should maybe not worry too much about trans-Planckian excursions of the inflaton field is based on the following logic.

 In order to generate
Planck suppressed higher-dimensional operator in the effective field theory, one has to  integrate out degrees of freedom. Apart from the
gravitons (more later), these might be 
heavy states (possibly with a bare Planckian mass). Consider, for instance, 
 fermions field coupled to the inflaton through a Yukawa coupling which gives them an extra mass  of the form $g\phi$, being $g$ a coupling constant. 
 
 If during inflation $\phi\gg M_p$, then these fermions will have trans-Planckian masses
 (unless $g$ is tiny). As  discussed in Ref. \cite{dvali},  trans-Planckian massive states 
do not describe independent quantum degrees of freedom,  but rather macroscopic classical states. The latter are then described by other 
 light fundamental degrees of freedom, such as the massless gravitons, and in fact are just classical black holes. In other words, it is possible that all states to which the inflaton is coupled to during the inflationary phase are classical black holes. 
If true, this fact immediately implies that
operators obtained by integrating out such trans-Planckian massive states will be exponentially suppressed at least by the Boltzmann factor $e^{-S}$, where $S \simeq g^2\phi^2//M_{p}^2$ is the Bekenstein-Hawking entropy. As a consequence, dangerous higher-dimensional operators of the form $( {\cal O}_n=\phi^n/M_{p}^{n-4})$ obtained after integrating out such trans-Planckian massive states are  Boltzmann suppressed and enter the effective Lagrangian as 

\be
{\cal L}\supset e^{-S} \frac{\phi^n}{M_p^{n-4}},
\ee
 nullifying in  this way all  potentially higher-dimensional operators 
 ${\cal O}_n$. 

%
%
Our final argument that trans-Planckian values of the inflaton might be harmless is based on the fact that the effective potential is actually an expansion in the tree-level potential and its second derivatives. Therefore, the anticipated expansion of the effective potential as  $V_{\rm eff}=V+\sum_n c_n \phi^n/M_p^{n-4}$  might be   reorganised and written  as an expansion in terms of $V$ and $V''$ (primes indicating here differentiation with respect to the inflation field) as these are the physical quantities corresponding to the energy density and mass squared respectively. 

This point is certainly not new and it was well-stressed, for example,  by Linde in Ref.  \cite{linde2} (see also Ref. \cite{ign}). 


%
To be more specific, let us consider a scalar $\phi$ coupled to gravity and write the metric fluctuations above a background with metric $\bar{g}_{\mu\nu}$ as
 \begin{eqnarray}
 g_{\mu\nu}=\bar{g}_{\mu\nu}+\kappa h_{\mu\nu},
 \end{eqnarray} 
 where 
 $\kappa^2=2/M_p^2$.
 Then, the quadratic part of the action turns out to be
\begin{align}
{\cal L}&=
 -\frac{1}{4} h_{\kappa\lambda}D^{\kappa\lambda,\rho\sigma}h_{\rho\sigma}
 -\frac{\kappa^2}{4}\left(2h^{\mu\lambda}h^{\nu}_\lambda-h^\kappa_\kappa h^{\mu\nu}\right)
 \partial_\mu \phi \partial_\nu \phi\nonumber \\
 &- \frac{\kappa^2}{16}\left(2 h^{\lambda\sigma}h_{\lambda\sigma}-(h_\nu^\nu)^2\right)(\partial_\mu\phi\partial^\mu \phi-V(\phi)),
 \end{align} 
where $D^{\mu\nu,\kappa\lambda}$ is an invertible differential operator. 
The one-loop effective potential will be given by integrating 
out the $h_{\mu\nu}$.
The result  in this case is (with some corrections with respect to Ref.   \cite{smolin})
\begin{align}
V_{\rm one-loop}&=\frac{1}{2}\ln \det\left(\Box-V''\right)\nonumber \\
&+\ln \det \left(D^{\mu\nu,\kappa\lambda} -I^{\mu\nu,\kappa\lambda}V\right),
\end{align}
where $I^{\mu\nu,\kappa\lambda}$ is a tensor constructed out of products of $\delta_\mu^\nu$'s. This is a formal expression for the effective potential and it needs to be regularized with a cut-off scale (of the order of $M_p$).
Using constant scalar field configurations, 
it is clear that after integrating out gravitons, the one-loop effective potential turns out to be a function of    $V$
and $V''$. Indeed, an explicit calculation \cite{smolin} reveals that the effective 
potential is of the form 

\begin{eqnarray}
  V_{\rm eff}=V(\phi)+ M_p^4 \sum_{nm} c_{nm} \frac{V''^nV^m}{M_p^{4(n+m)}},
  \end{eqnarray}  
when quantum gravity effects are taken into account. Again, there are nowhere 
$\phi^n/M_P^{n-4}$ terms 
and the inflationary predictions are not spoiled as long as $V\ll M_p^4$ and 
$V''\ll M_p^2$. 

It may also happens that due to an  undelrying symmetry, trans-Planckian   values of the fields are harmless. For example, it is known that string theory on compactified on a circle  of radius $R$ (or on spaces with $U(1)$ isometries in general) has the T-duality symmetry $R\to \ell_s^2/R$ where $\ell_s$ is the string scale, which interchanges winding and momentum states.  This symmetry has a fixed point at $R=\ell_s$. This implies that the theory defined  on a circle with   radius $R>\ell_s$ is actually identical to a circle of radius $R<\ell_s$ from the string point of view. One may try to implement the same idea for the inflaton itself by residing to a similar symmetry to restrict the possible values of the inflaton to sub-Planckian region. For example, one may assume that the inflaton is part 
 a complex field $\tau$ 
with a potential which is invariant under $SL(2,\mathbb{Z})$ transformations generated by  $\tau\to -M_p^2/\tau$  and $\tau\to \tau+M_p$ \cite{FLST}. In such a  case, if the inflaton for example was the modulus $|\tau|$, trans-Planckian values $|\tau|>M_p$ are 
equivalent to sub-Planckian values $|\tau|<M_p$. However, in this case one should also ensure that $SL(2,\mathbb{Z})$ is not broken by quantum gravity effects, although it is believed that in string theory such symmetries are indeed exact.

Let us also note that one can construct  class of single small field models of inflation that can predict, contrary to popular wisdom, an observable gravitational wave signal in the CMB  anisotropies \cite{bru}. Finally, it might well be that the observed tensor modes generated at the linear level are subdominant with respect to those created by a spectator scalar field with speed of sound lower than unity (in such a case the spectral index $n_T$ of the tensor
modes can be easily larger than zero)
\cite{mattei}.

For sure, the detection of a large level of tensor modes from inflation  will spur the community towards a better understanding of some crucial theoretical issues, possibly  with some interesting connections  to low-energy physics too.



\begin{references}


\bibitem{guth81} A. Guth, Phys. Rev. D {\bf 23}, 347 (1981)
\bibitem{lrreview} D. H. Lyth and A. Riotto, Phys.~Rept.~314 1 (1999);
A.~Riotto, hep-ph/0210162; W.~H.~Kinney,
astro-ph/0301448.
\bibitem{muk81} V. F. Mukhanov and G. V. Chibisov, JETP  Lett. {\bf 33}, 532 
(1981).
\bibitem{bardeen83} J. M. Bardeen, P. J. Steinhardt, and M. S. Turner, Phys. 
Rev. D {\bf 28}, 679 (1983).

\bibitem{bicep}  P.~A.~RAde {\it et al.}  [ BICEP2 Collaboration],
  arXiv:1403.3985 [astro-ph.CO].


\bibitem{ade}
  P.~A.~R.~Ade {\it et al.}  [Planck Collaboration],
  arXiv:1303.5082 [astro-ph.CO].



\bibitem{ng}  P.~A.~R.~Ade {\it et al.}  [Planck Collaboration],
  arXiv:1303.5084 [astro-ph.CO].

\bibitem{enq} K.~Enqvist and J.~Sirkka,
  Phys.\ Lett.\ B {\bf 314}, 298 (1993)
  [hep-ph/9304273].
  
\bibitem{noi1}  J.~Elias-Miro, J.~R.~Espinosa, G.~F.~Giudice, G.~Isidori, A.~Riotto and A.~Strumia,
  Phys.\ Lett.\ B {\bf 709}, 222 (2012)
  [arXiv:1112.3022 [hep-ph]].
  
  \bibitem{noi2} J.~R.~Espinosa, G.~F.~Giudice and A.~Riotto,
  JCAP {\bf 0805}, 002 (2008)
  [arXiv:0710.2484 [hep-ph]].

\bibitem{me} A. Riotto, 
  Phys.\ Rev.\ D {\bf 86}, 125038 (2012)
  [arXiv:1211.1321 [hep-ph]].

\bibitem{lythgrav} D.~H.~Lyth,
Phys.\ Rev.\ Lett.\  {\bf 78}, 1861 (1997).
  
  
  



\bibitem{curvaton1} K.~Enqvist and M.~S.~Sloth,
Nucl.\ Phys.\ B {\bf 626}, 395 (2002)
[arXiv:hep-ph/0109214].




\bibitem{LW}
D.~H.~Lyth and D.~Wands,
Phys.\ Lett.\ B {\bf 524}, 5 (2002).

\bibitem{curvaton3} T.~Moroi and T.~Takahashi,
Phys.\ Lett.\ B {\bf 522}, 215 (2001)
[Erratum-ibid.\ B {\bf 539}, 303 (2002)]
[arXiv:hep-ph/0110096].






\bibitem{LUW}
D.~H.~Lyth, C. Ungarelli and D.~Wands,  
Phys.~Rev.D {\bf 67}, 023503 (2003).

\bibitem{gamma1} G.~Dvali, A.~Gruzinov and M.~Zaldarriaga,
arXiv:astro-ph/0303591.

\bibitem{gamma2} 
L.~Kofman,
arXiv:astro-ph/0303614.

\bibitem{gamma3}
G.~Dvali, A.~Gruzinov and M.~Zaldarriaga,
  Phys.\ Rev.\ D {\bf 69}, 023505 (2004)
  [astro-ph/0303591].
  
  \bibitem{end1}
 D.~H.~Lyth,
  JCAP {\bf 0511}, 006 (2005)
  [astro-ph/0510443].
    
  \bibitem{end2} D.~H.~Lyth and A.~Riotto,
  Phys.\ Rev.\ Lett.\  {\bf 97}, 121301 (2006)
  [astro-ph/0607326].
  
%
%
%
%
%
%


\bibitem{reviewng} N.~Bartolo, E.~Komatsu, S.~Matarrese and A.~Riotto,
  Phys.\ Rept.\  {\bf 402}, 103 (2004)
  [astro-ph/0406398].


\bibitem{alberto} G.~Dvali and S.~Kachru,
  In *Shifman, M. (ed.) et al.: From fields to strings, vol. 2* 1131-1155
  [hep-th/0309095]; L.~Pilo, A.~Riotto and A.~Zaffaroni,
  JHEP {\bf 0407}, 052 (2004)
  [hep-th/0401004]; 
L.~Pilo, A.~Riotto and A.~Zaffaroni,
  Phys.\ Rev.\ Lett.\  {\bf 92}, 201303 (2004)
  [astro-ph/0401302].


\bibitem{dvali} G.~Dvali and C.~Gomez,
  arXiv:1005.3497 [hep-th].
 
\bibitem{linde2} 
  A.~D.~Linde,
  Contemp.\ Concepts Phys.\  {\bf 5}, 1 (1990)
  [hep-th/0503203].

\bibitem{ign} N.~Kaloper and L.~Sorbo,
  Phys.\ Rev.\ Lett.\  {\bf 102}, 121301 (2009)
  [arXiv:0811.1989 [hep-th]];
  N.~Kaloper, A.~Lawrence and L.~Sorbo,
  JCAP {\bf 1103}, 023 (2011)
  [arXiv:1101.0026 [hep-th]].
 
  \bibitem{smolin} L.~Smolin,
  Phys.\ Lett.\ B {\bf 93}, 95 (1980).

\bibitem{FLST} 
  S.~Ferrara, D.~Lust, A.~D.~Shapere and S.~Theisen,
  Phys.\ Lett.\ B {\bf 225}, 363 (1989).

\bibitem{bru} I.~Ben-Dayan and R.~Brustein,
  JCAP {\bf 1009}, 007 (2010)
  [arXiv:0907.2384 [astro-ph.CO]]; S.~Hotchkiss, A.~Mazumdar and S.~Nadathur,
  JCAP {\bf 1202}, 008 (2012)
  [arXiv:1110.5389 [astro-ph.CO]].
  
  
  \bibitem{mattei} M.~Biagetti, M.~Fasiello and A.~Riotto,
  Phys.\ Rev.\ D {\bf 88}, 103518 (2013)
  [arXiv:1305.7241 [astro-ph.CO]].
\end{references}
\end{document}